\documentstyle[12pt]{article}

\newcommand{\be}{\begin{equation}}
\newcommand{\ee}{\end{equation}}
\newcommand{\dd}{\partial}
\newcommand{\bea}{\begin{eqnarray}}
\newcommand{\eea}{\end{eqnarray}}

\begin{document}
\baselineskip .25in
\newcommand{\numero}{hep-th/9611218}  

\newcommand{\titre}{On the Eleventh Dimension of String Theory}
\newcommand{\auteura}{ } 
\newcommand{\auteurb}{Noureddine Mohammedi\,\,}

\newcommand{\placea}{Department of Physics,\\University of
Southampton,\\Southampton, SO17 1BJ, \\ U.K. }
\newcommand{\placeb}{Laboratoire de Math\'ematiques et Physique Th\'{e}orique
\footnote{UPRES A6083 du CNRS}\\
Universit\'e Fran\c{c}ois Rabelais\\
Facult\'e des Sciences et Techniques\\ Parc de Grandmont\\F-37200
 Tours - France.}
\newcommand{\beq}{\begin{equation}}
\newcommand{\eeq}{\end{equation}}

\newcommand{\abstrait}
{An explanation of the origin of the hidden eleventh dimension
in string theory is given. It is shown that any two sigma models 
describing the propagation of string backgrounds are related to 
each other by a Weyl transformation of the world-sheet metric. 
To avoid this ambiguity in defining two-dimensional sigma models, 
extra fields are needed. An interesting connection is established 
with Abelian T-duality.}
\begin{titlepage}
\hfill \numero  \\
\vspace{.5in}
\begin{center}
{\large{\bf \titre }}
\bigskip \\ \auteura 
\auteurb \footnote{e-mail:
nouri@celfi.phys.univ-tours.fr}
\bigskip \\ 
\placeb  \bigskip \\
\vspace{.9 in} 
{\bf Abstract}
\end{center}
\abstrait
 \bigskip \\
\end{titlepage}
\newpage
\section{Introduction}

There is by now an overwhelming evidence for a unifying origin
of all string theories. This ``grand unified theory'', known 
as M-theory or F-theory, is supposed to describe 
in a unified manner the non-perturbative r\'egime of
all five superstring theories
(See refs.\cite{review1,review2,review3} 
for some reviews and references
therein). It seems, therefore,
that the five different theories are just five different 
perturbative manifestations of the same underlying theory
\cite{witten1}. The 
emergence of M-theory or F-theory was inevitable 
after the discovery of S-duality in superstrings
\cite{ibanez,rey}. 
This duality, which is the
equivalent of electromagnetic duality in Yang-Mills theories,
relates two, \`a priori, different superstring theories
\cite{witten1,hull}. 
\par
The curious point about M-theory is that it lives in eleven
dimensions (or in twelve dimensions in the case of F-theory).
The various superstring theories are then obtained by a
variety of compactification procedures down to lower dimensions.
Given their common origin, one naturally expects to find relations
(usually refered to as duality transformations) between the resulting
theories. There are mathematical justifications for the apppearence of 
extra dimensions beyond ten; the critical dimension of superstrings.
Indeed, eleven dimensions is the maximum spacetime dimension in which
a consistent supersymmetric theory, containing no massless
particles with spins greater than two, can be constructed
\cite{nahm,julia}. It is also
the dimension, with one time direction,  in which supersymmetric 
extended object
(super $p$-branes) are naturally embedded 
\cite{hughes,bergshoeff,evan}. 
Allowing for more than one time 
direction is also another way of building super $p$-branes in more 
than eleven dimensions
\cite{12dim}. The latter construction is the essence of
F-theory \cite{vafa}.
\par 
It seems, therefore,  that there are hidden dimensions in superstring 
theories. Their origin is still,  however, unclear. It is easier to 
imagine a lower dimensional theory as descending from a higher
dimensional one but the reverse process is much less convincing.
An early attempt to explain this hidden dimension was put 
forward in \cite{duff1,townsend2,schmid,tseytlin1}. 
It relies on dualizing a vector field into a scalar
on the world-volume of a supermembrane propagating in ten 
dimensions. This method icreases the number of scalars by one.
We give, in this note, a natural explanation of the origin of
the hidden dimensions in superstrings. We examin this 
issue at the level of the two-dimensional sigma model and rely
on a crucial property of these models, namely their conformal 
invariance.
\par
A remarkable fact about M-theory is that it does not contain 
a dilaton field and treats equally all the massless modes 
of string theory \cite{witten1}. 
The dilaton field appears, as a component
of the eleven-dimensional metric, only
after dimensional reduction \cite{witten1,townsend}.
The expectation value of the dilaton field (and the other moduli
of the compactification)
plays the r\^ole of the small parameter of perturbation theory
in ordinary string theories. There are therefore no small
parameters in M-theory and hence the non-perturbative aspect
of this theory.
\par
In contrast, the dilaton field in a non-linear sigma model
is treated in a special manner. It is its coupling to the 
geometry of the two-dimensional world-sheet which distinguishes
it from the rest of the massless modes. This coupling is usually
given in the form \cite{fradkin}
\be
\int{\mathrm{d}}^2x\sqrt\gamma \Phi R^{(2)}\,\,\,,
\ee
where $\gamma_{\mu\nu}$ is the world-sheet metric, $\gamma$ is its
determinant and $R^{(2)}$ is its corresponding scalar curvature.
The coupling of the other massless modes involves the metric 
$\gamma_{\mu\nu}$ and not its derivatives. Furthermore, the dilaton 
term in a sigma model breaks conformal invariance at the classical
level. This last observation will be crucial to us here. It allows
for an ambiguity in defining sigma models in two dimensions. It will
be shown that the presence of the dilaton term makes it possible to 
connect any two models by a simple Weyl transformation of the 
metric $\gamma_{\mu\nu}$. We will show, in section two, that in 
order to evade this problem and to preserve conformal invariance
at the classical level, further fields must be introduced. These
will increase the dimension of the target spacetime. Finally
we comment, in section three, on a relation between these extra
fields and the fields used in the construction of Abelian 
T-duality (See \cite{duality} for a review on T-duality). 

\section{Weyl Transformations}

Our starting point is the two-dimensional theory defined by the 
action\footnote{We will deal only with bosonic theories here.
The supersymmetric case follows in a straightforward manner.}
\be
S = \int {\mathrm{d}}^2x L\left(x\right)
=\int {\mathrm{d}}^2x \left[{\mathcal{L}}\left(x\right)
+\sqrt\gamma \Phi R^{(2)}\right]\,\,\,,
\ee
where ${\mathcal{L}}$ is a two-dimensional Lagrangian 
and $\Phi\left(x\right)$ is the dilaton field of a string theory.
Let also ${\widetilde S}$ be another two-dimensional action given by
\be
{\widetilde S}=\int {\mathrm{d}}^2x {\widetilde L}\left(x\right)\,\,\,.
\ee
We would like to explore whether the action ${\widetilde S}$ is in anyway
related to the action $S$. We will indeed show that ${\widetilde S}$
can be obtained from $S$ by a Weyl rescaling of the metric $\gamma_{\mu\nu}$.
This is due, as we will see, to the presence of the dilaton field and 
to its particular coupling to the scalar curvature.
\par
In order to see this, we consider a local Weyl rescaling of the metric
\be
\gamma_{\mu\nu}\longrightarrow 
\exp\left[\sigma\left(x\right)\right] \gamma_{\mu\nu}\,\,\,.
\label{weyl}
\ee
In $d$-dimensions the transformation of the $d$-dimensional Ricci scalar
$R^{(d)}$ is given by\footnote{Our conventions are such that
$R^\mu_{\nu\rho\sigma}=\dd_\rho \Gamma^\mu_{\nu\sigma} +
\Gamma^\mu_{\rho\alpha}\Gamma^{\alpha}_{\nu\sigma} - \left(
\rho \leftrightarrow \sigma\right)$ and $R_{\mu\nu}
=R^\alpha_{\mu\alpha\nu}$.}
\be
R^{(d)}\longrightarrow 
\exp\left(-\sigma\right)
\left[R^{(d)} + \left(1-d\right)
\nabla^2\sigma +{1\over 4}\left(1-d\right)\left(d-2\right)
\gamma^{\mu\nu}\dd_\mu\sigma\dd_\nu\sigma\right]\,\,\,,
\ee 
where $\nabla^2$ is the usual curved Laplacian constructed from
the metric $\gamma_{\mu\nu}$. 
Specialising to  $d=2$ we see
that the last term in the transformation of $R^{(d)}$
vanishes and the Weyl transformation of the 
action $S$ is therefore given by
\bea
S &\longrightarrow& \int{\mathrm{d}}^2x \left[
{\mathcal{L}}\left(x\right) + 
\sqrt\gamma\Phi R^{(2)} -\sqrt\gamma\Phi\nabla^2\sigma\right]
\nonumber\\
&=& \int{\mathrm{d}}^2x\left[L\left(x\right) -\sqrt\gamma
\Phi\nabla^2\sigma\right]\,\,\,.
\eea
We assumed, for simplicity, that ${\mathcal{L}}$
is classically invariant under Weyl rescaling.
\par
Let now $G\left(x,y\right)$ denote the inverse of $\nabla^2$,
that is the Green's function defined by 
\be
\nabla^2_x G\left(x,y\right)={1\over\sqrt
{\gamma\left(x\right)}}
\delta^{(2)}\left(x-y\right)
\,\,\,,
\ee
where $\nabla^2_x$ is the Laplacian acting at the point $x$.
To get the two-dimensional action ${\widetilde S}$ from $S$
throught a Weyl transformation of the form 
(\ref{weyl}), it is sufficient
to choose the scale factor $\sigma$ as follows
\be
\sigma\left(x\right)= \int{\mathrm{d}}^2y G\left(x,y\right)
{1\over \Phi\left(y\right)}\left[L\left(y\right)
-{\widetilde L}\left(y\right)\right]\,\,\,.
\ee
Therefore, as long as the dilaton field is different from zero, 
any two-dimensional theory ${\widetilde S}$ can be obtained 
from $S$ by a Weyl transformation of the world-sheet metric
$\gamma_{\mu\nu}$.
\par
In the case when $S$ is a two-dimensional non-linear sigma 
model describing the propagation of string massless modes,
we have 
\be
\int{\mathrm{d}}^2x{\mathcal{L}}\left(x\right)=
\int{\mathrm{d}}^2x\left[\sqrt\gamma\gamma^{\mu\nu}
G_{ij}\left(X\right)\dd_\mu X^i\dd_\nu X^j 
+\epsilon^{\mu\nu}
B_{ij}\left(X\right)\dd_\mu X^i\dd_\nu X^j\right]\,\,\,,
\label{string}
\ee
where $G_{ij}$ and $B_{ij}$, ($i,j=1,\dots,D$),  
are the target space
metric and the antisymmetric tensor field respectively. 
This Lagrangian is
indeed classically invariant under a Weyl transformation. Therefore
any duality transformation relating two different string backgrounds
can be understood as a consequence of a Weyl transformation 
relating their corresponding sigma models.
\par
One way of avoiding the ambiguity in defining sigma models is to 
consider, instead of $S$, the following modified action
\be
S_{{\mathrm{mod}}}=\int{\mathrm{d}}^2x\left[
{\mathcal{L}}\left(x\right) +\sqrt\gamma\Phi R^{(2)}
+a\sqrt\gamma\gamma^{\mu\nu}\Phi\dd_\mu A_\nu
+b\sqrt\gamma\Phi \nabla^2 Y\right]\,\,\,,
\ee
where ${\mathcal{L}}$ is the sigma model Lagrangian in 
(\ref{string}) and $Y$ and $A_\mu$ are two new fields
transforming as
\be
A_\mu \longrightarrow A_\mu -\dd_\mu\sigma
\,\,\,\,,\,\,\,\,
Y\longrightarrow Y + \sigma\,\,\,.
\label{gaugetrans}
\ee
These last transformations cancel the Weyl transformation
of the dilaton term  
and renders the action classically invariant under Weyl
rescaling provided that the two constants $a$ and $b$
satisfy $b-a=1$.
\par
However, these new fields will impose, at the classical level,
a strong constraint on the dilaton field namely, 
$\dd_i\Phi\left(X\right)=0$.  In order to have a general
dilaton field we add to our modified action the following
general invariant Lagrangian
\bea
S_{{\mathrm{add}}}&=& \int {\mathrm{d}}^2x\left\{
\sqrt\gamma\gamma^{\mu\nu}\left[H\left(X\right)
D_\mu Y D_\nu Y + P_i\left(X\right)\dd_\mu X^i D_\nu Y
\right]\right.
\nonumber\\
&+& \left.\epsilon^{\mu\nu} \left[
Q_i\left(X\right)\dd_\mu X^i
D_\nu Y + N\left(X\right)\dd_\mu A_\nu
\right]\right\}\,\,\,,
\eea
where we have defined the invariant covariant derivative
$D_\mu =\dd_\mu Y + A_\mu$. Therefore the non-linear sigma 
model that one should start with is given by
\be
I\left(X,Y,A\right)= S_{\mathrm{mod}}+S_{\mathrm{add}}\,\,\,.
\label{totalaction}
\ee
In this last action the gauge field $A_\mu$ appears
at most quadratically and can be eliminated through its
equations of motion. This procedure leads to a sigma 
model with $D+1$ scalar fields. Therefore the requirement
that a sigma model is conformally invariant at the
classical level leads to an extension of the 
dimension of the target space.
\par 
Notice that we could have used a scalar field 
instead of the gauge field $A_\mu$. This is equivalent
to choosing $A_\mu =\dd_\mu V$, where $V$ is a scalar
field transforming as $V\longrightarrow V -\sigma$.
The important point here is that one needs at least
two fields in order to render the action conformally
invariant and at the same time to keep the dilaton
unconstrained. 
We will see in the rest of this note a nice relation between
this construction and Abelian T-duality.
  
\section{Duality and Classical Conformal Invariance}

Consider now a non-linear sigma model whose traget space coordinates
are denoted $\phi^a$ with $a=1,\dots,D$ and an action given by
\be
S\left(\phi\right)= \int{\mathrm{d}}^2x\left\{
\sqrt\gamma\gamma^{\mu\nu}G_{ab}\left(\phi\right)
\dd_\mu\phi^a\dd_\nu\phi^b
+\epsilon^{\mu\nu}B_{ab}\left(\phi\right)\dd_\mu\phi^a
\dd_\nu\phi^b +\sqrt\gamma\Phi\left(\phi\right)R^{(2)}\right\}
\,\,\,.
\ee
As explained above, this action can be related to any other
two-dimensional theory by a simple Weyl rescaling of the
metric $\gamma_{\mu\nu}$.  It is therefore
understood that a modification of this action on the 
steps of (\ref{totalaction}) is 
necessary. It will be shown 
shortly that this ambiguity can be also 
resolved if the metric $G_{ab}$
has an Abelian isometry. Let us suppose that this is indeed the case. 
We then split the coordinates $\phi^a$ as $\phi^a=\left(X^i,Y\right)$
with $i=1,\dots,D-1$. We assume that the isometry, in this new 
coordinate system, acts as a simple translation on the coordinate
$Y$ only. The non-linear sigma model is then given by
\bea
S\left(X,Y\right)&=& \int{\mathrm{d}}^2x\left\{
\sqrt\gamma\gamma^{\mu\nu}\left[
G_{ij}\left(X\right)\dd_\mu X^i
\dd_\nu X^j + H\left(X\right)\dd_\mu Y\dd_\nu Y 
+P_i\left(X\right)\dd_\mu X^i \dd_\nu Y\right]\right.
\nonumber\\
&+&\left.
\epsilon^{\mu\nu}\left[B_{ij}\left(X\right)\dd_\mu X^i
\dd_\nu X^j  + Q_i\left(X\right)\dd_\mu X^i \dd_\nu Y\right]
+\sqrt\gamma \Phi\left(X\right)R^{(2)}\right\}
\,\,\,.
\label{originalaction}
\eea
This action is invariant under the global
shift
$Y\longrightarrow Y + \sigma$. To construct the Abelian T-dual
of this action, the global isometry is gauged through the introduction
of an invariant covariant derivative $D_\mu Y=\dd_\mu Y + A_\mu$
with the  gauge field $A_\mu$ transforming
as $A_\mu \longrightarrow A_\mu -\dd_\mu \sigma$. In order to obtain 
a dual sigma model having the same number of degrees of freedom 
$(D$ coordinates), the field strength of the gauge field 
is constrained to vanish \cite{verlinde}. 
This is achieved by means of a Lagrange
multiplier field $Z$.  The action which leads to the dual sigma model 
is therefore written as
\bea
S\left(X,Y,Z,A\right)&=& \int{\mathrm{d}}^2x\left\{
\sqrt\gamma\gamma^{\mu\nu}\left[G_{ij}\left(X\right)\dd_\mu X^i
\dd_\nu X^j + H\left(X\right)D_\mu Y D_\nu Y 
+P_i\left(X\right)\dd_\mu X^i D_\nu Y\right]\right.
\nonumber\\
&+&\left.
\epsilon^{\mu\nu}\left[B_{ij}\left(X\right)\dd_\mu X^i
\dd_\nu X^j  + Q_i\left(X\right)\dd_\mu X^i D_\nu Y
+ Z\dd_\mu A_\nu\right]\right.
\nonumber\\
&+&\left.\sqrt\gamma \Phi\left(X\right)R^{(2)}\right\}
\,\,\,.
\label{gaugedaction}
\eea
As it is well-known, the integration over the Lagrange multiplier
leads to $A_\mu = \dd_\mu V$ and upon replacing in the gauged action
(\ref{gaugedaction}) 
we get our original sigma model (\ref{originalaction}) with 
the field $Y$ replaced by the shifted field $\widetilde Y =Y +V$.
On the other hand, keeping the Lagrange multiplier and eliminating
the gauge field leads to the dual theory. In doing so, the field
$Y$ completely disappears (without any use of gauge-fixing)
and the Lagrange multiplier 
$Z$ plays the r\^ole of this missing field. Hence the number of fields
is the same in the original and in the dual theories.
\par
However, the dual theory obtained in this way may as well be obtained, 
as explained above,  by a simple Weyl rescaling of the world-sheet metric
of the action (\ref{originalaction}).  Therefore in order to give
a sense (and not a mere Weyl rescaling) to this T-duality, 
the right action to consider is not $S\left(X,Y,Z,A\right)$ but a 
modified one given by
\be
I\left(X,Y,Z,A\right) = S\left(X,Y,Z,A\right) +
\int{\mathrm{d}}^2x \sqrt\gamma\left[
a\gamma^{\mu\nu}\Phi\left(X\right)\dd_\mu A_\nu
+b\Phi\left(X\right)\nabla^2Y\right]
\,\,\,\,.
\label{modifieddual}
\ee
This action is now invariant, when $b-a=1$, 
under the finite local transformations
\be
\gamma_{\mu\nu}\longrightarrow \exp\left(\sigma\right)
\gamma_{\mu\nu}\,\,\,,\,\,\,
Y\longrightarrow Y + \sigma\,\,\,,\,\,\,
A_\mu \longrightarrow A_\mu -\dd_\mu \sigma\,\,\,.
\ee
The action (\ref{modifieddual}) is exactly of the form of the 
action of the previous section given in (\ref{totalaction}) in which
the gauge field is restricted to be pure gauge and 
the Lagrange multiplier $Z$ is identified with $N\left(X\right)$.  
A classical elimination of the gauge field from the action
(\ref{modifieddual}) yields a sigma model defined on a
$D+1$  dimensional target space. This is because the 
field $Y$ does no longer disappear as it happened in the case 
of the action (\ref{gaugedaction}).

\section{Conclusions}

We proved in this paper that a two-dimensional non-linear
sigma model  
with a dilaton field is not uniquely defined. This is mainly
due to the breaking of classical conformal invariance by
the dilaton coupling. It is shown that any 
two-dimensional theory 
is obtainable from a sigma model with a dilaton field 
through a formal Weyl rescaling of the world-sheet metric.
\par
In order to avoid this ambiguity in defining sigma models, 
conformal invariance must be preserved at the classical level.
This is achieved by the introduction of a scalar and a gauge
field (though other choices of fields are not excluded). The 
introduction of these fields increases the dimension of the 
target spacetime. At first sight the choice of these two fields
seems arbitrary. However, their physical interpretation is much 
more natural in the context of Abelian T-duality.
\par
What remains to be explored here are the renormalisation 
properties of the modified non-linear sigma model constructed
in this note. The requirement that conformal invariance holds at
the quantum level leads to imposing some constraints on the 
string backgrounds. These conditions are in turn derived as equations 
of motion of a target space action which is the low energy theory
of M-theory. Indeed, imposing conformal invariance on the partition
function of the action $I\left(X,Y,Z,A\right)$ leads to the 
equation
\be
\langle T^\mu_\mu +\dd_\mu J^\mu + 
{\delta I\over \delta Y}\rangle=0\,\,\,.
\ee
The first term is the trace of the energy-momentum  
tensor defined as $T_{\mu\nu}=
{1\over\sqrt\gamma}{\delta I
\over\delta\gamma_{\mu\nu}}$.
In the absence of $Y$ and $A_\mu$ this term 
usualy leads to the vanishing of the beta-functions. 
The second term is the current corresponding to the gauge field 
and is given by $J^\mu={\delta I\over\delta A_\mu}$.
It is clear that in the presence of the fields
$Y$ and $A_\mu$, the beta-functions must be modified. 
This modification
depends also on whether the two fields $Y$ and $A_\mu$ are treated
as quantum or as background fields.
The quantum treatment of the model constructed in this note 
and other related topics will be considered elsewhere.


\end{document}